\UseRawInputEncoding
\documentclass[hidelinks,prb,aps,superscriptaddress,amssymb,amsmath,reprint,noeprint,floatfix]{revtex4-2}
\usepackage{natbib}
\usepackage{braket}
\usepackage{bm}
\usepackage{graphicx}
\usepackage{hyperref}
\usepackage{xcolor}
\hypersetup{bookmarksnumbered}

\newcommand{\FigOne}{
    \begin{figure}[t!]
        \centering
        \includegraphics[width=\linewidth]{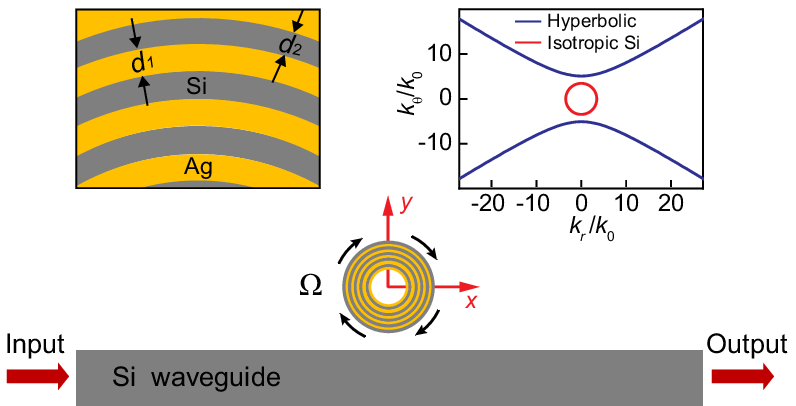}
        \caption{Schematic of the proposed optical isolator consisting of a subwavelength spinning particle near a waveguide. The left inset shows a zoom-in of the metamaterial. The right inset shows the equifrequency lines of the effective hyperbolic medium and isotropic silicon.}
        \label{fig:1}
    \end{figure}
}

\newcommand{\FigTwo}{
    \begin{figure}[t!]
        \centering
        \includegraphics[width=\linewidth]{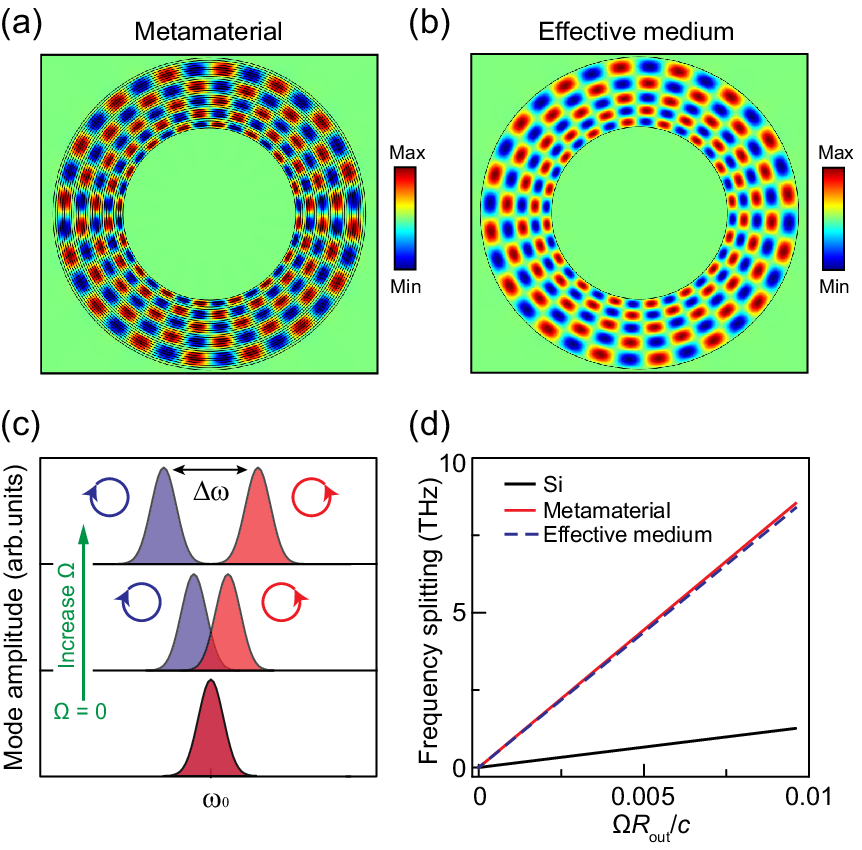}
        \caption{Chiral mode of (a) the hyperbolic metamaterial particle and (b) the effective-medium particle. (c) Schematic showing the frequency splitting induced by rotation. (d) Comparison of frequency splitting between the hyperbolic particle and a normal silicon particle.}
        \label{fig:2}
    \end{figure}
}

\newcommand{\FigThree}{
    \begin{figure}[t!]
        \centering
        \includegraphics[width=\linewidth]{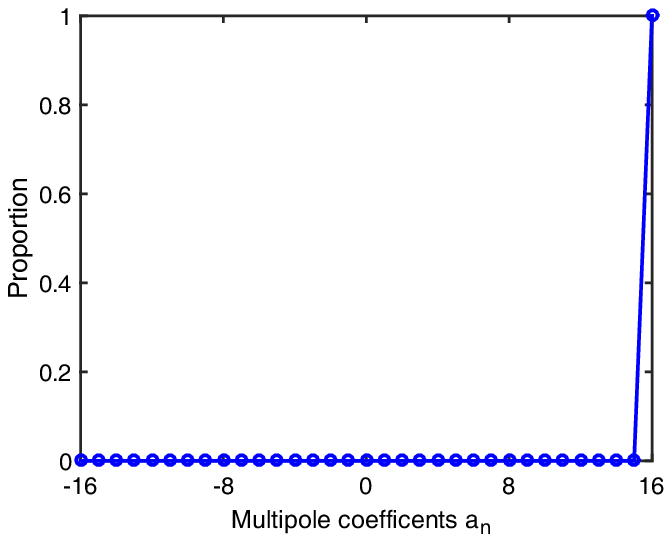}
        \caption{Multipole coefficient $a_n$ for the chiral mode in Fig. 2(a).}
        \label{fig:3}
    \end{figure}
}

\newcommand{\FigFour}{
    \begin{figure}[t!]
        \centering
        \includegraphics[width=\linewidth]{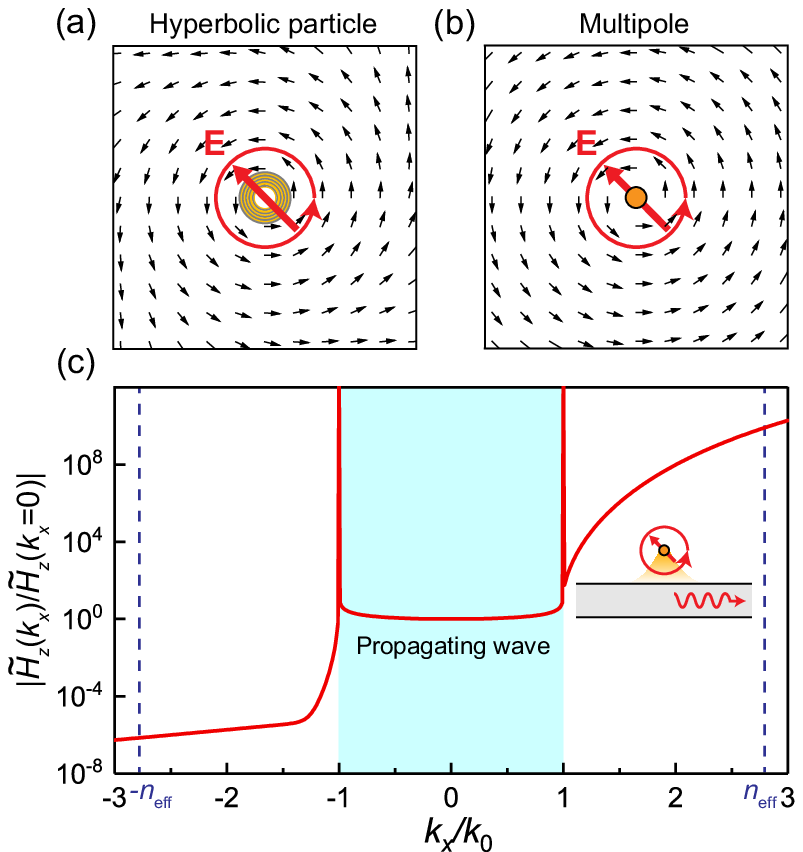}
        \caption{Poynting vectors in (a) the chiral mode of the hyperbolic particle and (b) the corresponding multipole with $m = 16$. The red arrows denote the average electric field vector, and the arrowed circles show the circulating direction of the electric field (i.e., spin). (c) Normalized spectral amplitude of the $H_z$ field for the multipole in panel (b). The inset schematically shows the unidirectional coupling under transverse spin-momentum locking.}
        \label{fig:4}
    \end{figure}
}

\newcommand{\FigFive}{
    \begin{figure}[t!]
        \centering
        \includegraphics[width=\linewidth]{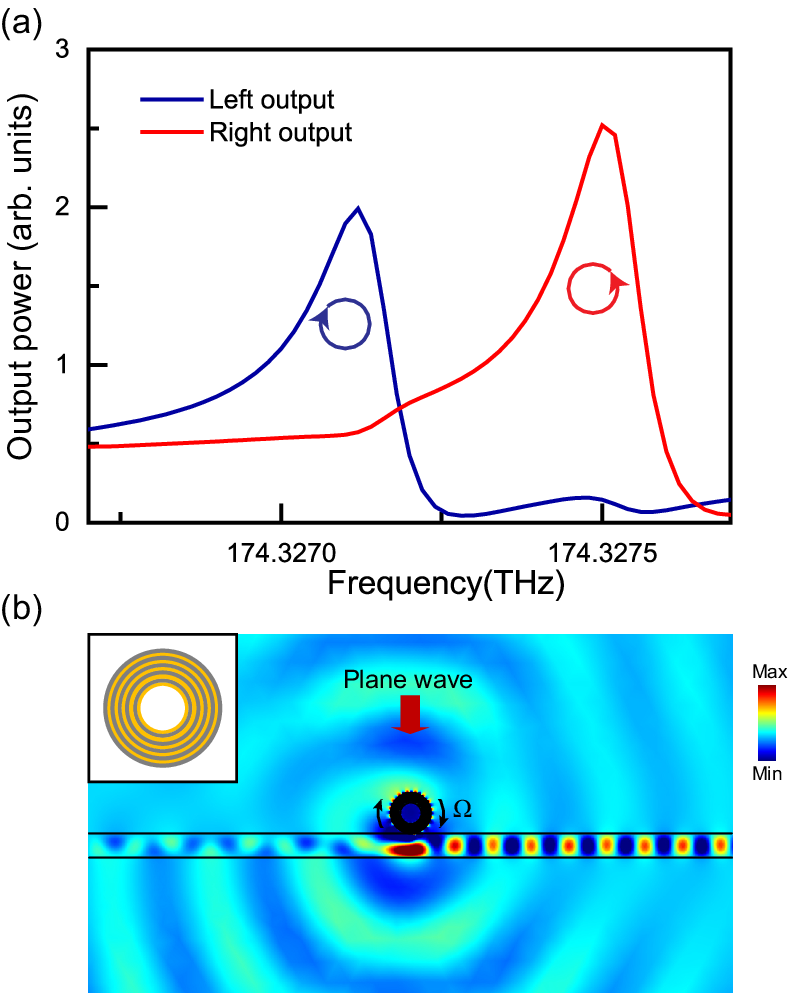}
        \caption{Output power at the left and right ends of the waveguide under plane wave excitation. (b) $H_z$ field at the second resonance. The background field due to the direct scattering of the waveguide is excluded. }
        \label{fig:5}
    \end{figure}
}

\newcommand{\FigSix}{
    \begin{figure}[t!]
        \centering
        \includegraphics[width=\linewidth]{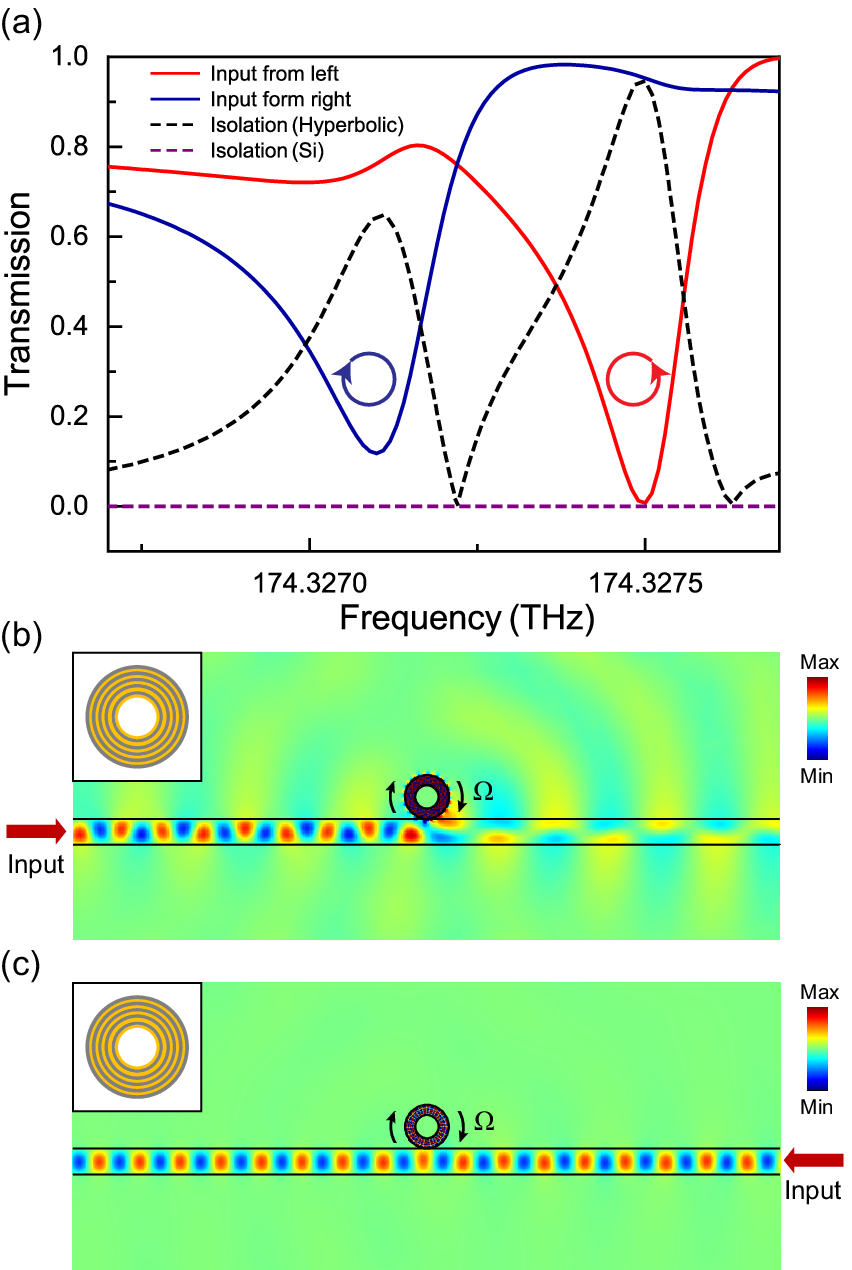}
        \caption{(a) Transmission under the excitation of left/right input of the waveguide. (b) $H_z$ field under the left input, showing a vanished transmission field. The residual field on the right side of the particle is attributed to the scattering of the particle. (c) $H_z$ field under the right input.}
        \label{fig:6}
    \end{figure}
}

\newcommand{\FigSeven}{
    \begin{figure}[t!]
        \centering
        \includegraphics[width=\linewidth]{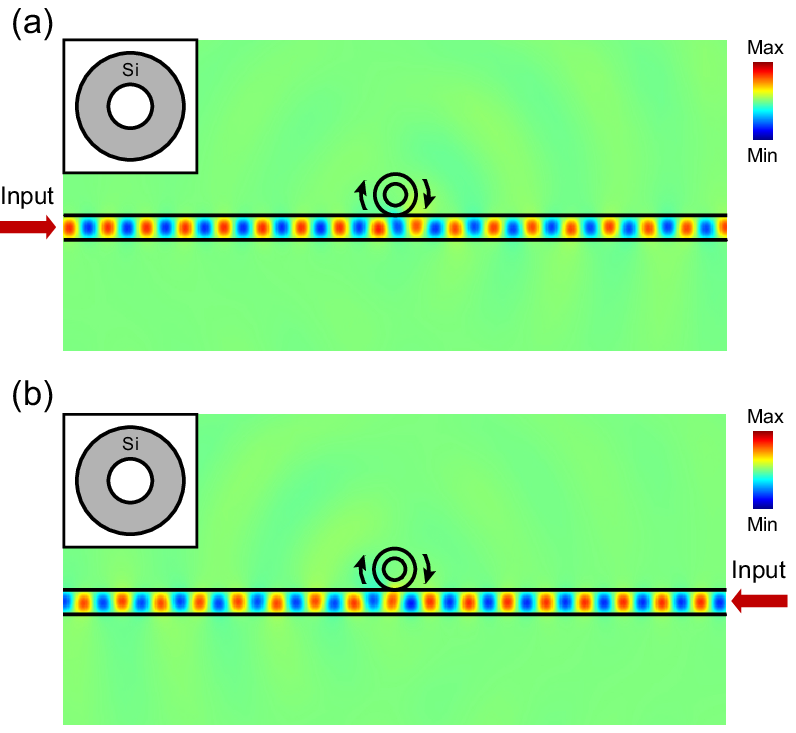}
        \caption{$H_z$ field for a spinning silicon particle under the  excitation at (a) the left input and (b) the right input of the waveguide.}
        \label{fig:7}
    \end{figure}
}

\begin{document}

\title{Optical isolation induced by subwavelength spinning particle via spin-orbit interaction}
\date{\today}

\author{Hongkang Shi}
\affiliation{School of Optical and Electronic Information, Huazhong University of Science and Technology, Wuhan 430074, China}
\affiliation{Department of Physics, City University of Hong Kong, Hong Kong, China}
\author{Yuqiong Cheng}
\affiliation{Department of Physics, City University of Hong Kong, Hong Kong, China}
\author{Zheng Yang}
\affiliation{Department of Physics, City University of Hong Kong, Hong Kong, China}
\author{Yuntian Chen}\email{yuntian@hust.edu.cn}
\affiliation{School of Optical and Electronic Information, Huazhong University of Science and Technology, Wuhan 430074, China}
\affiliation{Wuhan National Laboratory of Optoelectronics, Huazhong University of Science and Technology, Wuhan 430074, China}
\author{Shubo Wang}\email{shubwang@cityu.edu.hk}
\affiliation{Department of Physics, City University of Hong Kong, Hong Kong, China}
\affiliation{City University of Hong Kong Shenzhen Research Institute, Shenzhen, Guangdong 518057, China}

\begin{abstract}
Optical isolation enables nonreciprocal manipulations of light with broad applications in optical communications. Optical isolation by rotating structures has drawn considerable attention due to its magnetic-free nature and unprecedented performance. Conventional rotation-based optical isolation relies on the use of bulky cavities hindering applications in subwavelength photonics. Here, we propose a novel mechanism of optical isolation by integrating the unique dispersion of a hyperbolic metamaterial with the transverse spin-orbit interaction of evanescent waves. We show that rotation of a subwavelength hyperbolic nanoparticle breaks the time-reversal symmetry and yields two resonant chiral modes that selectively couple to the transverse spin of waveguide modes. Remarkably, the transverse spin-orbit interaction can give rise to unidirectional coupling and $>95\%$ isolation of infrared light at an experimentally feasible rotation speed. Our work fuses the two important fields of optical isolation and photonic spin-orbit interactions, leading to magnetic-free yet compact nonreciprocal devices for novel applications in optical communications, chiral quantum optics, and topological photonics.
\end{abstract}
\maketitle

\section{Introduction}
Nonreciprocity, i.e., an asymmetric response under interchange of the source and observation point, can give rise to one-way transport of light and thus optical isolation, which has numerous novel applications, such as invisible sensing \cite{dutton_understanding_1998} and noise-tolerant quantum computing \cite{sounas_non-reciprocal_2017,lodahl_chiral_2017}. Only a few approaches can be applied to achieve optical isolation, including nonlinearity \cite{fan_all-silicon_2012,peng_parity-time-symmetric_2014,chang_parity-time_2014}, spatiotemporal modulation \cite{yu_complete_2009,sounas_giant_2013,estep_magnetic-free_2014,Hafezi_opto_2012,shen_realization_2016,Seunghwi_dynamic_2019}, and external biasing \cite{casimir_onsagers_1945,metelmann_nonreciprocal_2015,fang_generalized_2017}. Nonlinear materials only work for high-intensity signals, while time modulation faces challenges in achieving fast and robust effects at optical frequencies \cite{sounas_non-reciprocal_2017}. The available external biasing technique routinely relies on a magnetic field, which is incompatible with on-chip optical devices due to the lack of magneto-optical materials at high frequencies. Alternatively, external biasing via structural rotation can be a promising solution to magnet-free nonreciprocal devices, as evident from the unprecedented performance of recent structural rotation-based optical isolators \cite{maayani_flying_2018,huang_nonreciprocal_2018,zhang_breaking_2020,jiao_2020}. However, these designs typically employ the whispering gallery modes of bulky cavities which are unsuitable for subwavelength photonics.

In this article, we propose a mechanism of optical isolation based on a spinning hyperbolic particle with a radius of $\sim\lambda / 5$ by taking advantage of the transverse spin-orbit interaction (SOI). Transverse SOI is a generic property of evanescent waves in various systems including waveguides \cite{petersen_chiral_2014,le_feber_nanophotonic_2015,young_polarization_2015,wang_arbitrary_2019} and metal surfaces \cite{rodriguez-fortuno_near-field_2013,bliokh_extraordinary_2014,wang_lateral_2014,oconnor_spinorbit_2014}. The intrinsic locking between the transverse spin angular momentum and the direction of linear momentum allows spin-dependent control of light propagation, leading to pseudo-nonreciprocial phenomena such as unidirectional coupling for one of the spin states \cite{lodahl_chiral_2017,wang_arbitrary_2019}. We show that with the particle rotation, a synthetic gauge field emerges and serves as a bias to remove the spin degeneracy of the chiral modes of the particle. Consequently, the transverse SOI in the particle-waveguide configuration can lead to truly unidirectional coupling beyond pseudo-nonreciprocity and thus optical isolation.

\textcolor{black}{The use of transverse SOI to achieve optical isolation at single-photon level has been experimentally realized in a coupled atom-waveguide configuration \cite{sayrin_nanophotonic_2015}, where the internal states of atoms with a biased magnetic field were employed to break the time-reversal symmetry and selectively couple with guided photons, instead of a macroscopic nonmagnetic spinning particle as we proposed here. This mechanism is suitable for optical isolation with low light levels due to the inherent saturation effect. In addition, the configuration of a spinning sphere interacting with a planar surface has been investigated for Casimir force physics, where it is shown that spinning induces asymmetric vacuum/thermal fluctuations around the sphere and gives rise to a lateral Casimir force \cite{manjavacas_lateral_2017}. In contrast to the aforementioned work, we will provide a different perspective and focus on the classical optical isolation properties of a spinning hyperbolic particle coupled with a dielectric waveguide.}

The paper is organized as follows. In Sec. II, we present the formulations of the wave equations describing the spinning hyperbolic particle with an effective-medium description and discuss the frequency splitting of the chiral modes. In Sec. III, we employ multipole expansions and angular spectrum analysis to understand the spin properties of the chiral modes and the associated transverse SOI. In Sec. IV, we numerically demonstrate the optical isolation phenomena induced by the spinning hyperbolic particle with comparison to a spinning silicon particle and present analytical formulations to understand the phenomena. We then discuss possible experimental realizations and draw conclusion in Sec. V.  

\FigOne

\section{The spinning hyperbolic particle}
 We consider a two-dimensional (2D) spinning cylindrical particle (angular velocity of $\Omega$) located above a silicon slab waveguide ($\varepsilon_{\text{wg}}=11.9$) with a distance of $g$, as shown in Fig. \ref{fig:1}. The particle has inner radius $R_{\text{in}}$ and outer radius $R_{\text{out}}$ and consists of multiple layers of silver and silicon with respective thicknesses of $d_1$ and $d_2$, as shown in the left inset in Fig. \ref{fig:1}. For simplicity, we take $d_1 = d_2 = d$ \textcolor{black}{and assume that the particle is empty in the center (corresponding to the white region)}. The relative permittivity of silver is described by a Drude model $\varepsilon_{\text{Ag}}=\varepsilon_{\infty}-\omega_{\text{p}}^2/(\omega^{2}+i\omega\gamma)$ with $\varepsilon_{\infty}=3.92$, $\omega_{\text{p}}=1.33 \times 10^{16}$ \textcolor{black}{rad/s}, and $\gamma=2.73 \times 10^{13}$ \textcolor{black}{rad/s} \cite{song_improving_2010}. We neglect material loss and consider only the real part of $\varepsilon_{\text{Ag}}$ in the numerical simulations. The multilayer structure is known as a hyperbolic metamaterial, which in the long-wavelength limit can be described as an effective homogeneous and anisotropic medium characterized by the dispersion relation $k_{\theta}^2/\varepsilon_r-k_r^2/\left| \varepsilon_{\theta} \right|=\omega^2/c^2$, where $\varepsilon_{\theta}=(\varepsilon_{\text{Si}}+\varepsilon_{\text{Ag}})/2$ and $\varepsilon_r=2\varepsilon_{\text{Si}}\varepsilon_{\text{Ag}}/(\varepsilon_{\text{Si}}+\varepsilon_{\text{Ag}})$ are the effective permittivities along principle axises \cite{poddubny_hyperbolic_2013}.  The right inset in Fig. \ref{fig:1} shows the equifrequency lines of the effective hyperbolic medium in comparison with that of isotropic silicon. The hyperbolic medium supports an unbounded wavevector due to its special dispersion relation \cite{yang_experimental_2012} \textcolor{black}{and can support resonance modes of the particle with large azimuthal quantum number. These modes have much larger frequency splitting compared with ordinary modes at the same rotational speed, which is essential in realizing the subwavelength optical isolator. We will elaborate on this point later in this section.}

Under the effective medium description, the electromagnetic properties of the spinning particle are governed by the Minkowski constitutive relations \cite{minkowski_notitle_1908}: 
\begin{equation}
\begin{aligned}
      &\mathbf{D}+\mathbf{v} \times \mathbf{H} / c^{2}=\boldsymbol{\varepsilon}(\mathbf{E}+\mathbf{v} \times \mathbf{B}),\\ 
      &\mathbf{B}+\mathbf{E} \times \mathbf{v} / c^{2}=\boldsymbol{\mu}(\mathbf{H}+\mathbf{D} \times \mathbf{v}),
\end{aligned}
\label{eq:1}
\end{equation}
 where $\boldsymbol{\varepsilon}$ and $\boldsymbol{\mu}$ are the effective permittivity and permeability tensors, $c$ is the speed of light, and $\mathbf{v}$ is the linear velocity of the material. For the considered  particle,  $\boldsymbol{\varepsilon}$ and $\boldsymbol{\mu}$ have only diagonal elements $\left\{\varepsilon_{r} \varepsilon_{0}, \varepsilon_{\theta} \varepsilon_{0}, \varepsilon_{z} \varepsilon_{0}\right\} \text { and }\left\{\mu_{0}, \mu_{0}, \mu_{0}\right\}$, respectively. Here, $\varepsilon_0$ and $\mu_0$ are the vacuum permittivity and permeability. The constitutive relations can be packed into matrix form: 
 \begin{equation}
\left[\begin{array}{l}
\mathbf{D} \\
\mathbf{B}
\end{array}\right]=\left[\begin{array}{cc}
{\boldsymbol{\varepsilon}}' & {\boldsymbol{\chi}}_{\text{em}} \\
{\boldsymbol{\chi}}_{\text{me}} & {\boldsymbol{\mu}}'
\end{array}\right]\left[\begin{array}{l}
\mathbf{E} \\
\mathbf{H}
\end{array}\right],
\label{eq:2}
\end{equation}
 which is the well-known constitutive relation for bi-anisotropic media; i.e., rotation transforms the original anisotropic medium into a bi-anisotropic medium. In conventional bi-anisotropic media with reciprocity, we have $ {\boldsymbol{\varepsilon}}'={\boldsymbol{\varepsilon}}'^{\mathrm{T}}, {\boldsymbol{\mu}}'={\boldsymbol{\mu}}'^{\mathrm{T}}, \text { and } {\boldsymbol{\chi}}_{\mathrm{em}}=\left({\boldsymbol{\chi}}_{\mathrm{me}}^{*}\right)^{\mathrm{T}}=-{\boldsymbol{\chi}}_{\mathrm{me}}^{\mathrm{T}}$, corresponding to Pasteur media \cite{sihvola_electromagnetic_1994}. Here “T” and “*” denote the transpose and the complex conjugate, respectively. However, for the spinning hyperbolic particle, we have ${\boldsymbol{\varepsilon}}'={\boldsymbol{\varepsilon}}'^{\mathrm{T}}, {\boldsymbol{\mu}}'={\boldsymbol{\mu}}'^{\mathrm{T}}, \text { and } {\boldsymbol{\chi}}_{\mathrm{em}}=\left({\boldsymbol{\chi}}_{\mathrm{me}}^{*}\right)^{\mathrm{T}}={\boldsymbol{\chi}}_{\mathrm{me}}^{\mathrm{T}}$ \cite{shi_gauge-field_2019}, corresponding to Tellegen media that break time-reversal symmetry and reciprocity \cite{sihvola_electromagnetic_1994}.

Under TM polarization, the constitutive relations can be manipulated into the following form
\begin{equation}
\left[\begin{array}{l}
D_{r} \\
D_{\theta} \\
B_{z}
\end{array}\right]=\left[\begin{array}{ccc}
\varepsilon_{r}^{\prime} & 0 & A_{\theta} / c \\
0 & \varepsilon_{\theta}^{\prime} & 0 \\
A_{\theta} / c & 0 & \mu^{\prime}
\end{array}\right]\left[\begin{array}{c}
E_{r} \\
E_{\theta} \\
H_{z}
\end{array}\right],   
\label{eq:3}
\end{equation}
where $\varepsilon_{r}^{\prime}=\varepsilon_{r}\left(\Lambda^{2}-1\right) /\left(\varepsilon_{r} \Lambda^{2}-1\right)$, $\varepsilon_{\theta}^{\prime}=\varepsilon_{\theta}$, $\mu^{\prime}=\left(\Lambda^{2}-1\right) /\left(\varepsilon_{r} \Lambda^{2}-1\right)$, $A_{\theta}=\Lambda\left(1-\varepsilon_{r}\right) /\left(\varepsilon_{r} \Lambda^{2}-1\right)$,  $\Lambda=r \Omega/c$ is the normalized rotation speed, and $k_0$ is the wavevector in vacuum \cite{shi_gauge-field_2019}. \textcolor{black}{Note that we use non-primed variables ($\varepsilon_r, \varepsilon_{\theta}, \varepsilon_{z}$, etc.) to denote the material properties of the stationary particle and primed variables ( $\varepsilon_{r}^{\prime}, \varepsilon_{\theta}^{\prime}, \mu^{\prime}$, etc.) to denote the material properties of the rotating particle.} Insert Eq. (\ref{eq:3}) into the source-free Maxwell’s equations with time harmonic dependence $e^{-i\omega t}$, we obtain
\begin{equation}
\left[\begin{array}{ccc}
-\frac{1}{r} \frac{\partial}{\partial \theta}-i k_{0} A_{\theta} & \frac{1}{r} \frac{\partial}{\partial r} r & -i k_{0} \mu^{\prime} \\
i k_{0} \varepsilon_{r}^{\prime} & 0 & \frac{1}{r} \frac{\partial}{\partial \theta}+i k_{0} A_{\theta} \\
0 & i k_{0} \varepsilon_{\theta}^{\prime} & -\frac{\partial}{\partial r}
\end{array}\right]\left[\begin{array}{c}
E_{r} \\
E_{\theta} \\
Z_{0} H_{z}
\end{array}\right]=0, 
\label{eq:4}
\end{equation}
where $k_0$ is wavevector in vacuum and $Z_0=\sqrt{\mu_0/\varepsilon_0}$ is the impedance of free space. Eliminate the electric field components $E_r,E_\theta$, we obtain 
\begin{equation}
\begin{split}
    \left(\frac{1}{r} \frac{\partial}{\partial \theta}+i k_{0} A_{\theta}\right)^2 H_{z}&+\frac{\varepsilon_{r}^{\prime}}{\varepsilon_{\theta}^{\prime}} \frac{1}{r} \frac{\partial}{\partial r}\left(r \frac{\partial}{\partial r} H_{z}\right)\\
    &+k_{0}^{2} \varepsilon_{r}^{\prime} \mu^{\prime} H_{z}=0,
\end{split}
\label{eq:5}
\end{equation}
Notably, an additional term $ik_0A_\theta$ emerges and corresponds to a synthetic gauge field $\mathbf{A}=A_{\theta}\hat{\theta}$ arising from rotation of the particle \textcolor{black}{(see the details in our previous work \cite{shi_gauge-field_2019}).}
\FigTwo
Equation (\ref{eq:5}) allows solutions of the form $H_{z}=H_{z}(r) e^{\pm i m \theta}$, where $m$ is a positive integer that denotes the azimuthal quantum number. The solutions with $\pm m$ represent a pair of chiral modes circulating in counterclockwise and clockwise directions. In the static limit (i.e., $\Omega=0$), the synthetic gauge field $\mathbf{A}$  vanishes, and the two modes are degenerate and orthogonal due to the time-reversal symmetry. Figure \ref{fig:2}(a) and \ref{fig:2}(b) show the chiral mode with $m=+16$ in the static metamaterial particle and the corresponding effective-medium particle at $f=174.33$ THz, which are obtained via full-wave simulations using COMSOL \cite{COMSOL}. We set $R_{\text{in}}=185$ nm, $R_{\text{out}}=335$ nm, and $d=5$ nm (corresponding to a size of $R_{\text{out}}\approx \lambda / 5$). We note that the $H_z$ fields of the two systems agree well, demonstrating the validity of the effective-medium description.

At nonzero rotation speed (i.e., $\Omega \neq 0$), the synthetic gauge field $\mathbf{A}$ breaks the time-reversal symmetry and removes the degeneracy of the chiral modes, which can be considered a photonic analog of the Zeeman effect. This leads to a frequency splitting $\Delta \omega$ of the two modes. Substituting $H_{z}(r) e^{\pm i m \theta}=H_{z}(r) e^{\pm i k_{\theta} r \theta}$ into Eq. (\ref{eq:5}), we obtain 
\begin{equation}
    \frac{1}{r} \frac{\partial}{\partial r}\left[r \frac{\partial}{\partial r} H_{z}(r)\right]=\frac{\varepsilon_{r}^{\prime}}{\varepsilon'_{\theta}}\left[\left(k_{\theta} \pm k_{0} A_{\theta}\right)^{2}-k_{0}^{2} \varepsilon_{r}^{\prime} \mu^{\prime}\right] H_{z}(r),
    \label{eq:6}
\end{equation}
 which is an eigen equation with eigenvalues $\lambda=\varepsilon_{r}^{\prime}/\varepsilon'_{\theta}\left[k_{0}^{2}\left(A_{\theta}^{2}-\varepsilon_{r}^{\prime} \mu^{\prime}\right) \pm 2 k_{\theta} A_{\theta} k_{0}+k_{\theta}^{2}\right]$. The two eigenvalues give two solutions of $k_0$ which we denote as $k_1$ and $k_2$. Using Eq. (\ref{eq:6}) it is easy to find $\left(A_{\theta}^{2}-\varepsilon_{r}^{\prime} \mu^{\prime}\right)\left(k_{2}-k_{1}\right)=2 k_{\theta} A_{\theta}$, and the frequency splitting between the two modes is given by
\begin{equation}
    \Delta \omega=|\Delta k| c=\frac{2 m A_{\theta} c / r}{\varepsilon_{r}^{\prime} \mu^{\prime}-A_{\theta}^{2}}.
    \label{eq:7}
\end{equation}
In the limit of $\Lambda=R_{\text{out}} \Omega/c \ll 1$, all high-order terms about $\Lambda$ can be neglected, and the vector potential is reduced to $\mathbf{A}=\Lambda(\varepsilon_r-1)\hat{\theta}$.  Thus, the frequency splitting becomes
\begin{equation}
    \Delta \omega \approx 2 m \Omega\left(1-\frac{1}{\varepsilon_{r}}\right).
    \label{eq:8}
\end{equation}
Evidently, $\Delta \omega$ is proportional to the rotation angular velocity $\Omega$. Figure \ref{fig:2}(c) schematically shows the evolution of the chiral modes as $\Omega$ increases. The arrowed circles indicate the circulating direction of the mode field pattern. The two modes have identical spectra at $\Omega=0$ but are separated at $\Omega\neq 0$. In addition, $\Delta \omega$ is proportional to the azimuthal quantum number $m$. Thus, modes with larger $m$ have larger $\Delta \omega$ at the same rotation speed, which is critical to the realization of optical isolation. Such modes are absent in conventional subwavelength optical structures due to low dielectric constants. Fortunately, they can emerge in the hyperbolic particle due to the unbounded wavevector. To see the large frequency splitting of the hyperbolic particle, we compare it with that of a silicon particle, both of which share the same geometric parameters and rotation speed. At $f=152$ THz, the silicon particle supports a pair of magnetic dipole resonances, while the hyperbolic particle supports a pair of chiral modes with $m=\pm 8$. Their frequency splittings are shown in Fig. \ref{fig:2}(d) as the solid black and red lines, respectively, which are numerically computed using COMSOL. Evidently, the hyperbolic particle yields much larger frequency splitting due to the larger value of $m$. The dashed blue line denotes the analytical result based on effective-medium description, which agrees well with the numerical result of the metamaterial particle. We note that in the case of Fig. \ref{fig:2}(a) with $m=\pm 16$, the frequency splitting of the hyperbolic particle is even larger, while the counterpart modes are missing in the silicon particle due to cutoff frequencies.

\section{Angular spectrum analysis and transverse spin-orbit interaction}
To understand the scattering properties of the hyperbolic particle, we apply multipole expansions to its scattering field. Adopting the notations of Bohren and Huffman \cite{bohren_absorption_1998}, the radiation field of the chiral modes can be expressed as
\begin{equation}
\begin{aligned}
\mathbf{E}_{\mathrm{s}} &=-\sum_{n=-\infty}^{+\infty} E_{n}\left[b_{n} \mathbf{N}_{n}+i a_{n} \mathbf{M}_{n}\right], \\
\mathbf{H}_{s} &=\frac{i k}{\omega \mu} \sum_{n=-\infty}^{+\infty} E_{n}\left[b_{n} \mathbf{M}_{n}+i a_{n} \mathbf{N}_{n}\right],
\end{aligned}
\label{eq:9}
\end{equation}
where $E_n=E_0(-i)^n / k$ and $\mathbf{M}_n,\mathbf{N}_n$ are the vector harmonics defined in cylindrical coordinate system as:
\begin{equation}
\begin{aligned}
&\mathbf{M}_{n}=i n \frac{Z_{n}(k r)}{r} e^{i n \theta} \hat{r}-k Z_{n}^{\prime}(k r) e^{i n \theta} \hat{\theta}, \\
&\mathbf{N}_{n}=k Z_{n}(k r) e^{i n \theta} \hat{z}.  
\end{aligned} 
\label{eq:10}
\end{equation}
Here $Z_n=\mathcal{H}_n^{(1)}$ is the Hankel function of the first kind. In the considered system, we have $k=k_0,\mu=\mu_0$, and $n=m$. Since $\mathbf{H}=H_z \hat{z}$, the multipole coefficients $b_n$ vanishes. The coefficient $a_n$ can be numerically determined via the following integral:
\begin{equation}
    a_{n}=\frac{\int_{0}^{2 \pi} \mathbf{M}_{n}^{*} \cdot \mathbf{E}_{s} d \theta}{-i E_{n} \int_{0}^{2 \pi} \mathbf{M}_{n}^{*} \cdot \mathbf{M}_{n} d \theta},
    \label{eq:11}
\end{equation}
where “*” denotes the complex conjugate. Using the above equations, we found that the dominating coefficient for the chiral mode in Fig. 2(a) is $a_{16}$, as shown in Fig. \ref{fig:3}. Thus the $H_z$ field of the mode can be expressed as
\FigThree
\begin{equation}
    H_{z}=\frac{-k}{\omega \mu} a_{n} \mathcal{H}_{n}^{(1)}(k r) e^{i n \theta}, 
    \label{eq:12}
\end{equation}
where $n=m=16$. Thus, the hyperbolic particle can be considered a passive “point” multipole with $m=16$ due to its subwavelength nature (i.e., $R_{\text{out}}\approx \lambda / 5$).  Importantly, the multipole carries spin in the $z$ direction (i.e., transverse spin), which can be characterized by the Stokes parameter $S_3$ of the average electric field. The sign of $S_3$ has a one-to-one correspondence with the circulating direction of chiral modes, i.e., $S_3=1$ ($S_3=-1$) for the mode with $m>0$ ($m<0$), and thus can be used to label the chiral modes' electric field $\mathbf{E}_{\mathrm{p}}^{\pm}$ and the eigenfrequencies $\omega_{\pm}$. Figure \ref{fig:4}(a) and \ref{fig:4}(b) show the spin and Poynting vectors of $\mathbf{E}_{\mathrm{p}}^{\pm}$ and the corresponding multipole, respectively. The red arrows denote the average electric field of the particle, and the arrowed circles show the circulating direction of the electric field (i.e., spin), which is the same as the circulating direction of the mode field pattern in Fig. \ref{fig:2}(a). Notably, the Poynting vectors (black arrows) also circulate counterclockwisely, which indicates asymmetric propagation of the near field. This can be understood with the \textcolor{black}{2D} angular spectrum representation of $H_z$ field of the multipole: \cite{borghi_angular-spectrum_2004,novotny_principles_2012}: 
\FigFour
\FigFive
\begin{equation}
    H_{z}(x, y)=\int_{-\infty}^{\infty} \widetilde{H}_{z}\left(k_{x}, y\right) e^{i k_{x} x} \mathrm{d} k_{x},
    \label{eq:13}
\end{equation}
where $\widetilde{H}_z (k_x,y)$ is the spectrum amplitude that can be determined as
\begin{equation}
    \widetilde{H}_{z}\left(k_{x}, y\right)=\frac{1}{2 \pi} \int_{-\infty}^{\infty} H_{z}(x, y) e^{-i k_{x} x} \mathrm{d} x,
    \label{eq:14}
\end{equation}
For the spectral component at the position of the waveguide ($y=-g$), the normalized amplitude $|\widetilde{H}_{z}\left(k_{x}\right) / \widetilde{H}_{z}\left(k_{x}=0\right)|$ with $k_{x} \in\left[-3 k_{0}, 3 k_{0}\right]$ is shown in Fig. \ref{fig:4}(c) as the solid red line. The shaded region corresponds to the propagating waves with $|k_x / k_0| \leq 1$. For evanescent waves with $|k_x / k_0| > 1$, the spectral amplitude unambiguously shows asymmetry with respect to $+k_x$ and $-k_x$. \textcolor{black}{The coupling between the particle and the waveguide is determined by the overlap of their eigen fields, which requires matching of the wavevector $k_x$ of the component plane waves with the wavevector $k_{\text{eff}}=n_{\text{eff}} k_0$ of the guided wave (see Appendix A), as dictated by the momentum conservation \cite{picardi_unidirectional_2017,vazquez-lozano_near-field_2019}}. The effective refractive index of the fundamental TM mode of the waveguide is $n_{\text{eff}}=2.77$, as marked by the dashed blue lines in Fig. \ref{fig:4}(c). The spectral amplitude with $k_x=n_{\text{eff}} k_0$ is more than ten orders of magnitude larger than that of $k_x=-n_{\text{eff}} k_0$. Thus, the evanescent waves of the multipole couple unidirectionally to the waveguide mode, as shown by the inset in Fig. \ref{fig:4}(c), indicating locking between the transverse spin and the direction of linear momentum $k_x$. The unidirectional coupling can be verified by exciting the particle with a normally incident plane wave and calculating the output intensity at both ends of the waveguide. As shown in Fig. \ref{fig:5}(a), at the resonance frequency $\omega_{\pm}$, the output intensity at only one end dominates, depending on the spin of the chiral modes. Figure \ref{fig:5}(b) shows the magnetic field at the second resonance, which clearly shows the unidirectional coupling from the particle to the right side of the waveguide. Notably, such high-efficiency and spatially broadband near-field unidirectionality can only be achieved with high-order multipoles \cite{vazquez-lozano_near-field_2019}. It serves as another critical element of optical isolation in our settings.

\section{Optical isolation} 
We now consider the particle-waveguide configuration in Fig. \ref{fig:1} and demonstrate the optical isolation. The waveguide thickness is 400 nm. The particle locates at $g = 338$ nm above the waveguide and has a spinning frequency of $\Omega/(2\pi)=0.19$ GHz. The transmission of the fundamental TM waveguide mode is shown in Fig. \ref{fig:6}(a), where the solid red (blue) line denotes the excitation from the left (right) input. Both transmissions have a dip corresponding to one of the chiral modes, and the two dips are spectrally separated due to the frequency splitting effect. The arrowed circles denote the spin of the chiral modes. The large transmission contrast between the ``left" and ``right" excitations unambiguously shows the nonreciprocity of the system, which can be applied to realize optical isolation. The isolation ratio (i.e., the absolute difference between the blue and red lines) is shown as the dashed black line in Fig. \ref{fig:6}(a). Remarkably,  large isolation ratios occur at the two eigenfrequencies of the chiral modes (i.e., $\omega_{\pm}$). In particular, an isolation ratio of $>95\%$ is achieved at $\omega_{+}$. Figure \ref{fig:6}(b) and \ref{fig:6}(c) show the $H_z$ field under opposite excitations at $\omega_{+}$. As seen, light is almost completely blocked for excitation from the left input (the field outside the waveguide is attributed to the scattering of the particle), while it can largely propagate through the waveguide when excited from the right input. In addition, the optical isolation is robust against variations of the geometric parameters. Another design with $R_{\text{in}}=370$ nm, $R_{\text{out}}=670$ nm and $d=10$ nm can give $>94\%$ optical isolation at $f=150.30$ THz ($R_{\text{out}}\approx\lambda / 3$) (results not shown). For comparison, we also calculated the transmission properties of a normal silicon particle with the same size and spinning frequency, and the corresponding isolation is shown as dashed purple line in Fig. \ref{fig:6}(a). It is noticed that the isolation is negligible due to the absence of any resonance in the considered subwavelength regime. Figure \ref{fig:7} shows the magnetic field of the silicon-particle system at the same frequency of $\omega_{+}$. We see that light can largely propagate through the waveguide under the excitation at either input. 
\FigSix
\FigSeven

The optical isolation can be understood as a result of mode couplings under transverse spin-orbit interaction mediated by the spinning particle. \textcolor{black}{The input guided electric field can be expressed as
$\mathbf{E}_{\text {in }}=\mathbf{E}_{\mathrm{wg}}^{\pm}(y) e^{\pm i k_{\text {eff }}x}$, where $\mathbf{E}_{\mathrm{wg}}^{\pm}(y)$ characterize the $y$ dependence of the electric field and the superscript ``$\pm$" denotes the spin carried by the electric field. Note that the sign of spin is locked to the wave vector $\pm k_\text{eff}$. The output electric field is given by the following Lippmann-Schwinger-type equation (see Appendix A):
\begin{equation}
\mathbf{E}_{\mathrm{out}}=\mathbf{E}_{\mathrm{in}}+\frac{\omega^{2}}{\left(\omega_{\pm}\right)^{2}-\omega^{2}} i |\kappa_{\pm}|^{2} \mathbf{E}_{\mathrm{out}},
\label{eq:15}
\end{equation}
from which we obtain the transmitted field 
\begin{equation}
\mathbf{E}_{\text {out }}=\left[1+\frac{i\left|\kappa_{\pm}\right|^{2}}{\left(\omega_{\pm} / \omega\right)^{2}-1-i\left|\kappa_{\pm}\right|^{2}}\right] \mathbf{E}_{\text {in }}.
\label{eq:16}
\end{equation}
 Here $\kappa_{\pm}=2 \pi a\left(\varepsilon_{\mathrm{wg}}-1\right)\left(\mathbf{E}_{\mathrm{wg}}^{\pm}\right)^{*} \tilde{\mathbf{E}}_{\mathrm{p}}^{\pm}\left(\pm k_{\mathrm{eff}}\right)$ is the coefficient of coupling between the chiral modes and the guided modes, where $a$ is the effective width of waveguide, $\tilde{\mathbf{E}}_{\mathrm{p}}^{\pm}\left(\pm k_{\mathrm{eff}}\right)$ denotes the Fourier amplitude of $\mathbf{E}_{\mathrm{p}}^{\pm}$ for $k=\pm k_\text{eff}$ at $y=-g$, which is proportional to $\tilde{H}_z(\pm k_\text{eff})$ in Eq. (\ref{eq:14}). Note that material loss and radiation loss in the free space are neglected in the above formulations. Equation (\ref{eq:16}) indicates that the output field is attributed to the interference of the incident field and the reaction field due to the particle. At the resonance frequency $\omega=\omega_{\pm}$, the output field vanishes due to destructive interference. The transmission coefficient can be expressed as
\begin{equation}
t_{\pm}=\left|\frac{\mathbf{E}_{\text {out }}}{\mathbf{E}_{\text {in }}}\right|^{2}=1-\frac{|\kappa_{\pm}|^{4}}{\left|\left(\omega_{\pm} / \omega\right)^{2}-1\right|^{2}+|\kappa_{\pm}|^{4}}.
\label{eq:17}
\end{equation}
The isolation ratio $I=\left|t_{+}-t_{-}\right|$ can be obtained as
\begin{equation}
I=\left|\frac{|\kappa_{+}|^{4}}{|\left(\omega_{+} / \omega\right)^{2}-1|^{2}+|\kappa_{+}|^{4}}-\frac{|\kappa_{-}|^{4}}{|\left(\omega_{-} / \omega\right)^{2}-1|^{2}+|\kappa_{-}|^{4}}\right|.
\label{eq:18}
\end{equation}
The isolation ratio has two peak values at the resonance frequency $\omega=\omega_\text{p}^{+}$ and $\omega=\omega_\text{p}^{-}$, respectively, in accordance with the fullwave numerical results in Fig. \ref{fig:6}(a). Note that $I\left(\omega_{+}\right) \neq I\left(\omega_{-}\right)$ because rotation breaks the mirror symmetry and leads to $|\kappa_{+}| \neq |\kappa_{-}|$.}

\section{Discussion and conclusion}
Optical isolation by rotating structures relies on three critical elements: high rotation speed, resonance modes with a large azimuthal quantum number and a high Q factor. The three elements normally compromise each other: a high rotation speed can only be achieved for small structures which normally do not support the required resonance modes in the subwavelength regime. Thus, conventional rotation-based optical isolators typically have large volumes. Our system shoots three birds with one stone: the hyperbolic particle is geometrically small ($R_{\text{out}}\approx \lambda / 5$) and supports chiral modes with a large azimuthal quantum number and a high Q factor. 

\textcolor{black}{To realize the proposed optical isolator, a few challenges have to be addressed. The nonreciprocity derives from the relativistic effect induced by the particle spinning, thus, a high spinning frequency is required due to the large speed of light. To achieve the expected performance here, a spinning frequency of 0.19 GHz is desired. While this may not be realistic for conventional approaches, recent experiments have demonstrated GHz spinning of nanoparticles by using the optical tweezers technique \cite{reimann_ghz_2018,ahn_optically_2018,ahn_ultrasensitive_2020}, where a high vacuum environment combined with the approach of active parametric feedback could reduce the mechanical instability of the particle \cite{gieseler_subkelvin_2012}. Another challenge is the requirement of high-precision nanofabrication techniques. The hyperbolic particle consists of thin layers of metal-dielectric materials on the nanoscale. To guarantee the validity of the effective-medium description, fabrication imperfections must be smaller than the wavelength inside the hyperbolic structure. Such structures have been successfully fabricated and applied to achieve various optical functionalities \cite{schwaiger_rolled-up_2009,wang_metaparticles_2018,lin_nanoscale_2019}. Last but not least, the proposed isolator faces the common issue of loss as in all metal-based structures, which can reduce the Q factor of the resonances and undermine its performance. One possible solution is to add gain into the material, which would unavoidably add complication to the setup \cite{rong_all-silicon_2005,xiao_loss-free_2010,stockman_nanoplasmonics_2011,khurgin_how_2015,lin_nanoscale_2019}. The proposed structure can be modified to improve the experimental feasibility. For example, one may replace the silver-silicon hyperbolic layers by other metal-dielectric layers that are easier to fabricate, in which our discussions equally apply, as the mechanism relies on the hyperbolic dispersion of the corresponding effective medium. Besides, one can replace the air core in the particle center by a suitable dielectric core with a low refractive index.} 

As a concluding remark, optical isolation can be achieved with the spinning hyperbolic particle coupled to a waveguide via the transverse spin-orbit interaction. Our study thus paves a critical step towards on-chip optical isolation and may trigger further explorations of structural rotation-induced effects and phenomena, such as the nonreciprocal hopping in periodic structures and the non-Hermiticity due to asymmetric couplings. In addition, the small size of the optical insulator is perfect for integration with compact optical circuits, which can generate novel applications in optical communications, chiral quantum optics, and topological photonics.

\vspace{20pt}
\section*{acknowledgements}
The work described in this paper was supported by grants from the Research Grants Council of the Hong Kong Special Administrative Region, China (Project No. CityU 11301820) and the National Natural Science Foundation of China (No. 11904306 and No. 11874026).

\appendix
\section{Mode coupling under transverse spin-orbit interaction}
\textcolor{black}{
The response of the particle can be characterized by its Green's function \cite{novotny_principles_2012}
\begin{equation}
\mathbf{G}\left(\mathbf{r}, \mathbf{r}^{\prime}\right)=\sum_{n} \frac{c^2\mathbf{E}_{\mathrm{p}}^{n}(\mathbf{r})\left[\mathbf{E}_{\mathrm{p}}^{n}\left(\mathbf{r}^{\prime}\right)\right]^{*}}{\omega_{n}^{2}-\omega^{2}}
\label{eq:a1}
\end{equation}
which has been expanded in terms of the normalized eigenmodes $\mathbf{E}_p^n$ of the particle at eigenfrequency $\omega_n$. Here we have neglected the material loss and the radiation loss in free space. Under the excitation of external field $\mathbf{E}$, the particle's field can be obtained as
\begin{equation}
    \begin{aligned}
\mathbf{E}_{\text{p}}&=i \omega \mu_{0} \int \mathbf{G}\left(\mathbf{r}, \mathbf{r}^{\prime}\right) \cdot \mathbf{J}\left(\mathbf{r}^{\prime}\right) d A_{\text{p}}\\
&=\sum_{n} \frac{\omega^{2}}{\omega_{n}^{2}-\omega^{2}}\left[\int\left(\mathbf{E}_{\mathrm{p}}^{n}\right)^{*} \cdot \Delta \boldsymbol{\varepsilon} \cdot \mathbf{E} d A_{\text{p}}\right] \mathbf{E}_{\text{p}}^n
\end{aligned}
     \label{eq:a2}
\end{equation}
Where $\mathbf{J}=\partial \mathbf{P} / \partial t=-i \omega \varepsilon_{0} \Delta \boldsymbol{\varepsilon} \cdot \mathbf{E}$ is the polarization current, $\Delta\boldsymbol{\varepsilon}$ is the relative permittivity contrast between the structure and the background medium (i.e. free space), and the integration is evaluated over the volume of the particle $A_p$. Thus, the coefficient for the coupling from the guided mode to the particle mode is
\begin{equation}
\kappa_{\mathrm{pw}}=\int\left(\mathbf{E}_{\mathrm{p}}^{n}\right)^{*} \cdot \Delta \boldsymbol{\varepsilon} \cdot \mathbf{E}_{\mathrm{wg}} d A_{\text{p}}
\label{eq:a3}
\end{equation}
and the coefficient for the coupling from the particle mode to the guided mode is
\begin{equation}
\kappa_{\mathrm{wp}}=\int\left(\mathbf{E}_{\mathrm{wg}}\right)^{*} \cdot \Delta \boldsymbol{\varepsilon} \cdot \mathbf{E}_{\mathrm{p}}^{n} d A_{\mathrm{wg}}.
\label{eq:a4}
\end{equation}
Here $\mathbf{E}_{\mathrm{wg}}=\mathbf{E}_{\mathrm{wg}}^{\pm}(y) e^{\pm i k_{\mathrm{eff}} x}$ is the normalized guided mode with $\mathbf{E}_{\mathrm{wg}}^{\pm}(y)$ characterizing the $y$ dependence of the field, and the superscript ``$\pm$" denotes the spin of the field that is locked to the effective wavevector $\pm k_{\mathrm{eff}}$ of the guided mode. The two coefficients satisfy $\kappa_{\text{wp}}=\kappa_{\text{pw}}^*$ due to energy conservation \cite{haus_waves_1983,suh_temporal_2004}.  Let $\kappa_{\text{wp}}=\kappa$ and use the properties of Dirac delta function \cite{mathews_mathematical_1970}, we obtain
\begin{widetext}
\begin{equation}
\begin{aligned} 
\kappa=&\left(\varepsilon_{\mathrm{wg}}-1\right) \int\left[\mathbf{E}_{\mathrm{wg}}^{\pm}(y) e^{\pm i k_{\mathrm{eff}} x}\right]^{*} \mathbf{E}_{\mathrm{p}}^{n} d A_{\mathrm{wg}} \\
=&\left(\varepsilon_{\mathrm{wg}}-1\right) \iint\left[\mathbf{E}_{\mathrm{wg}}^{\pm}(y)\right]^{*} e^{\mp i k_{\mathrm{eff}} x}\left[\int \tilde{\mathbf{E}}_{\mathrm{p}}^{n}\left(k_{x}, y\right) e^{i k_{x} x} d k_{x}\right] d x d y \\
= & 2\pi\left(\varepsilon_{\mathrm{wg}}-1\right) \int\left[\mathbf{E}_{\mathrm{wg}}^{\pm}(y)\right]^{*}\left[\int \tilde{\mathbf{E}}_{\mathrm{p}}^{n}\left(k_{x}, y\right) \delta\left(k_x \mp k_{\mathrm{eff}}\right) d k_{x}\right] d y \\
= &2 \pi\left(\varepsilon_{\mathrm{wg}}-1\right) \int\left[\mathbf{E}_{\mathrm{wg}}^{\pm}(y)\right]^{*} \tilde{\mathbf{E}}_{\mathrm{p}}^{n}\left(\pm k_{\mathrm{eff}}, y\right) d y 
\approx  2 \pi a\left(\varepsilon_{\mathrm{wg}}-1\right)\left[\mathbf{E}_{\mathrm{wg}}^{\pm}(-g)\right]^{*} \tilde{\mathbf{E}}_{\mathrm{p}}^{n}\left(\pm k_{\mathrm{eff}},-g\right), \end{aligned}
\label{eq:a5}
\end{equation}
\end{widetext}
where the integrand has been evaluated at the upper surface of the waveguide $y=-g$ since the coupling is mainly contributed by the evanescent waves, $a$ is the effective width of the waveguide, $\tilde{\mathbf{E}}_{\mathrm{p}}^{n}$ denotes the Fourier amplitude of the field $\mathbf{E}_{\mathrm{p}}^{n}$. In the following, we omit the ``$-g$" argument for simplicity. In the considered frequency regime, the response of the particle is dominated by two chiral modes $\mathbf{E}_\text{p}^+$ and $\mathbf{E}_\text{p}^-$. According to the angular spectrum analysis in Sec. III, we have $\tilde{\mathbf{E}}_{\mathrm{p}}^{+}\left(+k_{\mathrm{eff}}\right) \gg \tilde{\mathbf{E}}_{\mathrm{p}}^{+}\left(-k_{\mathrm{eff}}\right)$ and $\tilde{\mathbf{E}}_{\mathrm{p}}^{-}\left(-k_{\mathrm{eff}}\right) \gg \tilde{\mathbf{E}}_{\mathrm{p}}^{-}\left(+k_{\mathrm{eff}}\right)$ (the Fourier amplitude of electric field is proportional to that of magnetic field). Thus, the following two coefficients are sufficient to characterize the coupling between the particle and the waveguide: 
\begin{equation}
\begin{array}{l}
\kappa_{+}=2 \pi a\left(\varepsilon_{\mathrm{wg}}-1\right)\left(\mathbf{E}_{\mathrm{wg}}^{+}\right)^{*} \tilde{\mathbf{E}}_{\mathrm{p}}^{+}\left(+k_{\mathrm{eff}}\right), \\
\kappa_{-}=2 \pi a\left(\varepsilon_{\mathrm{wg}}-1\right)\left(\mathbf{E}_{\mathrm{wg}}^{-}\right)^{*} \tilde{\mathbf{E}}_{\mathrm{p}}^{-}\left(-k_{\mathrm{eff}}\right)
\end{array}
\label{eq:a6}
\end{equation}
For input field $\mathbf{E}_{\text {in }}$, the output field  $\mathbf{E}_{\text {out}}$ is determined by the interference of the input field and the reaction field due to the particle. Using Eqs. (\ref{eq:a2})-(\ref{eq:a6}), we obtain the following Lippmann-Schwinger-type equation \cite{fan_theoretical_1999,sakurai_modern_2010}
\begin{equation}
\mathbf{E}_{\mathrm{out}}=\mathbf{E}_{\mathrm{in}}+\frac{\omega^{2}}{\left(\omega_{\pm}\right)^{2}-\omega^{2}} i |\kappa_{\pm}|^{2} \mathbf{E}_{\mathrm{out}},
\label{eq:a7}
\end{equation}
from which we obtain the output field 
\begin{equation}
\mathbf{E}_{\text {out }}=\left[1+\frac{i\left|\kappa_{\pm}\right|^{2}}{(\omega_{\pm}/\omega)^2-1-i\left|\kappa_{\pm}\right|^{2}}\right] \mathbf{E}_{\text {in }}.
\label{eq:8}
\end{equation}
The transmission coefficient is
\begin{equation}
t_{\pm}=\left|\frac{\mathbf{E}_{\text {out }}}{\mathbf{E}_{\text {in }}}\right|^{2}=1-\frac{|\kappa_{\pm}|^{4}}{\left|\left(\omega_{\pm} / \omega\right)^{2}-1\right|^{2}+|\kappa_{\pm}|^{4}}.
\label{eq:a9}
\end{equation}
The isolation ratio $I=\left|t_{+}-t_{-}\right|$ can be obtained as 
\begin{equation}
I=\left|\frac{|\kappa_{+}|^{4}}{|\left(\omega_{+} / \omega\right)^{2}-1|^{2}+|\kappa_{+}|^{4}}-\frac{|\kappa_{-}|^{4}}{|\left(\omega_{-} / \omega\right)^{2}-1|^{2}+|\kappa_{-}|^{4}}\right|,
\label{eq:a10}
\end{equation}
which gives rise to two resonance peaks at the resonance frequency $\omega=\omega_{+}$ and $\omega=\omega_{-}$, respectively. The peak values will reduce in the case with damping due to material loss or radiation of the particle. 
}

\bibliography{Reference}
\end{document}